\documentclass[aps,prl,twocolumn,superscriptaddress,showpacs,intlimits,amsmath,amssymb,floats]{revtex4-1}
\usepackage{bm,graphicx,epsfig}
\usepackage{gensymb}
\usepackage{wasysym}

\begin{document}

\title{The many faces of quantum kagome materials:\\
Interplay of further-neighbour exchange and Dzyaloshinskii-Moriya interaction}

\author{Tsezar F. Seman}
\affiliation{Department of Physics, Northern Illinois University, DeKalb, Illinois 60115, USA}
\affiliation{Advanced Photon Source, Argonne National Laboratory, Argonne, Illinois 60439, USA}

\author{Cheng-Chien Chen}
\affiliation{Advanced Photon Source, Argonne National Laboratory, Argonne, Illinois 60439, USA}

\author{Rajiv R. P. Singh}
\affiliation{Department of Physics, University of California, Davis, California 95616, USA}

\author{Michel van Veenendaal}
\affiliation{Department of Physics, Northern Illinois University, DeKalb, Illinois 60115, USA}
\affiliation{Advanced Photon Source, Argonne National Laboratory, Argonne, Illinois 60439, USA}

\begin{abstract}
The field of frustrated magnetism has been enriched significantly by the discovery of various kagome lattice compounds.
These materials exhibit a great variety of macroscopic behaviours ranging from magnetic orders to quantum spin liquids.
Using large-scale exact diagonalization, we construct the phase diagram of the $S=1/2$ $J_1$-$J_2$ kagome Heisenberg model with $z$-axis Dzyaloshinskii-Moriya interaction $D_z$.
We show that this model can systematically account for many of the experimentally observed phases.
Small $J_2$ and $D_z$ can stabilize respectively a gapped and a gapless spin liquid.
When $J_2$ or $D_z$ is substantial, the ground state develops a $\mathbf{Q}=0$, 120\degree antiferromagnetic order.
The critical strengths for inducing magnetic transition are $D^c_z \sim 0.1\, J_1$ at $J_2=0$, and  $J^c_2\sim 0.4\, J_1$ at $D_z=0$.
The previously reported values of $D_z$ and $J_2$ for herbertsmithite [ZnCu$_3$(OH)$_6$Cl$_2$] place the compound in close proximity to a quantum critical point.
\end{abstract}


\date{\today}

\maketitle

In frustrated magnetism~\cite{Ramirez_ARMS_1994,Diep2005, Lacroix2011}, the kagome lattice has become the paradigmatic system of choice for studying novel spin-liquid phases that result from geometric frustration and quantum fluctuation~\cite{Balents_Nature_2010}.
For example, the $S=1/2$ antiferromagnetic (AFM) kagome Heisenberg model with only the nearest-neighbour (NN) exchange $J_1$ is magnetically disordered with a spin correlation length less than one lattice spacing~\cite{Zeng_PRB_1990, Sachdev_PRB_1992, Leung_PRB_1993, Nakamura_PRB_1995, Lecheminant_PRB_1997, Jiang_PRL_2008,Yan_Science_2011,Depenbrock_PRL_2012}.
The ground state is characterized as a gapped topological spin liquid with a finite triplet gap $0.05-0.15\, J_1$~\cite{Waldtmann_EPJB_1998, Li_PRB_2007, Sing_PRB_2008, Jiang_PRL_2008, Lauchli_PRB_2011, Yan_Science_2011,Depenbrock_PRL_2012, Nishimoto_NatComm_2013} and a nonzero topological entanglement entropy~\cite{Jiang_NatPhys_2012, Depenbrock_PRL_2012}.
Advances in computational methods and theoretical ideas have led to a deeper understanding and classification of these exotic states of matter, which do not break any symmetry and sustain emergent fractional excitations.
However, a wide gulf still separates theory and experiments.

The celebrated herbertsmithite, ZnCu$_3$(OH)$_6$Cl$_2$~\cite{Braithwaite_MM_2004, Shores_JACS_2005}, has nearly perfect kagome planes consisting of $S$=1/2, Cu$^{2+}$ atoms.
Despite a predominant $J_1\sim 200$ K, this compound does not develop any long-range magnetic order down to $T= 50$ mK~\cite{Ofer_arXiv_2006, Mendels_PRL_2007, Helton_PRL_2007, Lee_NatMater_2007, Han_Nature_2012},
agreeing with the NN AFM kagome Heisenberg model.
But there are disagreements:
Without finding any apparent gap down to $0.1$ meV, neutron scattering suggests gapless excitations~\cite{Helton_PRL_2007,Han_Nature_2012}.
Magnetic susceptibility $\chi_m$ shows an upturn at low temperature~\cite{Ofer_arXiv_2006, Helton_PRL_2007},
which is also unexpected; a gapped spin liquid would otherwise show a vanishingly small $\chi_m$ close to $T=0$.
To account for these discrepancies, Dzyaloshinskii-Moriya (DM) interaction~\cite{Rigol_PRL_2007, Rigol_PRB_2007,Zorko_PRL_2008, Cepas_PRB_2008,Rousochatzakis_PRB_2009}, exchange anisotropy~\cite{Ofer_PRB_2009, Han_PRL_2012}, or quenched site dilution~\cite{Dommange_PRB_2003,Rigol_PRB_2007, Rozenberg_PRB_2008,Rousochatzakis_PRB_2009} have been investigated for herbertsmithite.
On the other hand, not all kagome compounds are spin liquids.
In materials such as Cu(1,3-bdc) [Cu-benzenedicarboxylate]~\cite{Nytko_JACS_2008,Marcipar_PRB_2009}, vesignieite [BaCu$_3$(VO$_4$)$_2$(OH)$_2$]~\cite{Colman_PRB_2011,Quilliam_PRB_2011,Yoshida_JMC_2012, Yoshida_JPSJ_2013}, and Cs$_2$Cu$_3$SnF$_{12}$~\cite{Ono_PRB_2009,Matan_NatPhysics_2010, Katayama_arXiv_2014},
the ground states is a $\mathbf{Q}=0$, 120\degree AFM order,
which highlights potential interactions beyond the NN exchange.
Determining the importance of additional couplings is thereby key to a comprehensive understanding of the diversified properties in kagome materials~\cite{Iqbal_arXiv_2015}.

In this Letter, we study the effects of the $z$-axis DM interaction $D_z$ and second NN exchange coupling $J_2$,
which are arguably the two most relevant perturbations for isotropic kagome compounds.
For the first time, we construct the phase diagram of the $J_1$-$J_2$-$D_z$ model using large-scale exact diagonalization (ED) with cluster sizes up to $N=42$ sites.
By also examining the excitation gaps and static structure factors, we show that the model can sustain various phases including long-range AFM order and quantum spin liquids with or without a finite spin gap.
The distinct ground states among different kagome compounds can be systematically accounted for with varying interaction strengths of the systems.
In particular, the reported values  of $D_z$ and $J_2$ for herbertsmithite indicate a ground state closely proximal to a magnetic quantum critical point, where small extra perturbations suffice to suppress its long-range magnetism.
Based on a numerically unbiased method, our study provides a direct road map for gauging $D_z$ and $J_2$ in $S=1/2$ materials with nearly prefect isotropic kagome structures.

{\it Model and Method --}
We consider the $J_1$-$J_2$ kagome Heisenberg model with DM interactions:
\begin{eqnarray}\label{eq:hamiltonian}
H= J_1\!\! \sum_{ <ij>} \! \mathbf{S}_i \cdot \mathbf{S}_j + J_2\!\!  \sum_{\ll ij \gg} \! \mathbf{S}_i \cdot \mathbf{S}_j + \sum_{<ij>} \! \mathbf{D}_{ij} \cdot (\mathbf{S}_i \times \mathbf{S}_j),\nonumber\\
\end{eqnarray}
where the first two terms are respectively superexchange interactions between NN and second NN $S=1/2$ sites.
The third DM-interaction term originates from relativistic spin-orbit coupling and is nonzero when lattice inversion symmetry is absent~\cite{Dzyaloshinsky_JPCS_1958, Moriya_PR_1960}.
Here we focus on the $z$-axis component of the DM interaction $\mathbf{D}_{ij} = D_z \mathbf{z}$, using the convention that $D_z >0$ when all links $i\rightarrow j$ are oriented clockwise [see inset of Fig. 1(a)]~\cite{Hiroki_JPSJ_2011}.
We neglect the in-plane component $D_\parallel$, as it is reported to be smaller than $D_z$ in materials of interests and also reducible to second order in $D^2_\parallel/J_1$ with a spin basis rotation~\cite{Shekhtman_PRL_1992, Cheng_PRB_2007, Cepas_PRB_2008}.
Throughout the paper we consider antiferromagnetic couplings ($J_1, J_2 > 0$ ) and set $J_1\equiv 1$.

We solve Eq. (\ref{eq:hamiltonian}) systematically by numerical diagonalization on clusters of size $N$ up to $N=42$.
The Hamiltonian matrix is constructed utilizing translational symmetry and $S^z_{\textrm{total}}\equiv \sum_i S^z_i$ conservation~\cite{Sandvik_AIPConfProc_2010}.
The resulting sparse matrix eigenvalue problem is solved by the Krylov-Schur algorithm as implemented in SLEPc~\cite{SLEPC} and PETSc~\cite{PETSC} libraries.
The cluster choices and calculation details are given in the {\it Supplemental Material}.

{\it Phase Diagram --}
We first establish the phase diagram of the $J_1$-$J_2$-$D_z$ model.
Without $D_z$ and $J_2$, the system is magnetically disordered.
A finite $D_z$ or $J_2$ could support a $\mathbf{Q}=0$, 120\degree AFM ground state with spins lying in the $xy$-plane [see inset of Fig. 1(a)].
Therefore, we proceed to map out the phase boundary between the quantum AFM state and the magnetically disordered region by studying the transverse spin-spin correlation function~\cite{Cepas_PRB_2008}:
\begin{eqnarray}\label{eq:sab}
S^{xx}_{ab}(\mathbf{Q}) \equiv \frac{24}{N^2} \sum_{IJ} e^{i\mathbf{Q}\cdot(\mathbf{R}_I - \mathbf{R}_J )} \langle S^x_{Ia} S^x_{Jb}\rangle.
\end{eqnarray}
Here, $\mathbf{R}_{I,J}$ are unit-cell positions and $a,b$ are intra-unit-cell site indices.
$S^{xx}_{ab}(\mathbf{Q})$ represents the elements of a $3\times3$ matrix and peaks at $\mathbf{Q}=0$.
The largest matrix eigenvalue $\equiv S^{xx}_{120\degree}(N)$ corresponds to the 120\degree AFM spin arrangements,
and its classical value is equal to 1 with the pre-factor choice of Eq. (\ref{eq:sab})~\cite{Cepas_PRB_2008}.
Spontaneous spin symmetry breaking can be identified on finite-size clusters by a linear $1/\sqrt{N}$ extrapolation~\cite{Huse_PRB_1988, Neuberger_PRB_1989, Sandvik_AIPConfProc_2010}.
When the extrapolated $S^{xx}_{120\degree} \equiv S^{xx}_{120\degree} (N=\infty)> 0$, long-range magnetic order develops.

\begin{figure}[t!]
\includegraphics[width=\columnwidth]{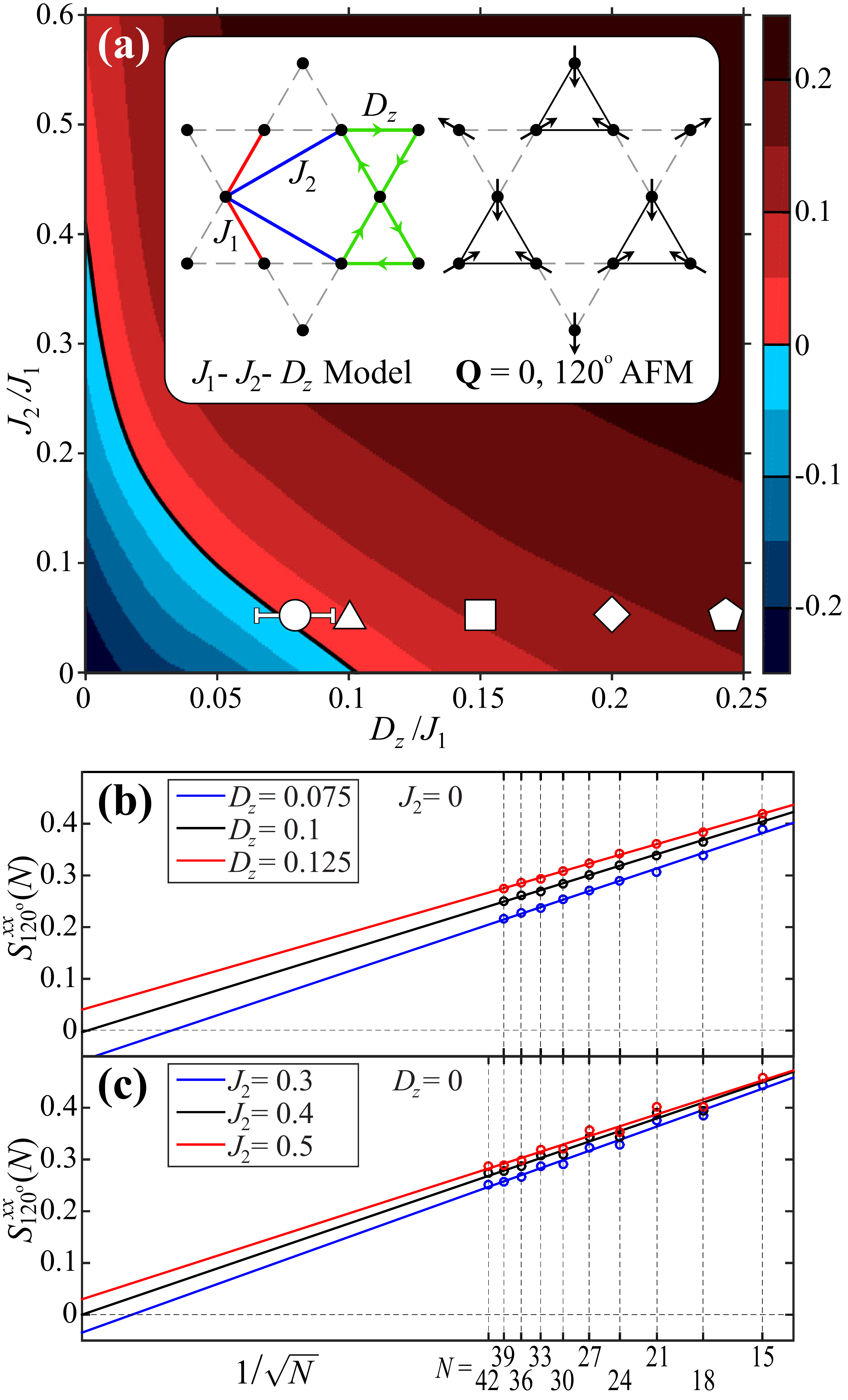}
\caption{
(a) Phase diagram of the $J_1$-$J_2$-$D_z$ model.
The false-color intensity represents $S^{xx}_{120\degree}$.
The system is magnetically disordered in the blue region, and it develops the $\mathbf{Q}=0$, 120\degree AFM order in the red.
The geometric symbols denote estimated values of $D_z$ in various kagome compounds:
$\Circle$ for herbertsmithite, $\vartriangle$ for Cu(1,3-bdc)$, \Square$ for vesignieite, $\Diamond$ for barlowite, and $\pentagon$ for Cs$_2$Cu$_3$SnF$_{12}$.
(b)-(c) Linear $1/\sqrt{N}$ extrapolations of $S^{xx}_{120\degree}(N)$ at various $D_z$ and $J_2$.
}\label{fig:sab}
\end{figure}

Figure 1(a) shows the phase diagram obtained by a grid interpolation of $11\times 7 =77$ points on the $(D_z,J_2)$ plane.
The generic features are computed with cluster sizes $N=15 - 33$, and further $N=36,\, 39$ calculations are performed to more precisely locate the phase boundary.
The blue region of Fig. 1(a) represents a magnetically disordered ground state ($S^{xx}_{120\degree} < 0$), and the red represents the $\mathbf{Q}=0$, 120\degree AFM phase ($S^{xx}_{120\degree}  > 0$).
Figures 1(b) and 1(c) show linear $1/\sqrt{N}$ extrapolations of $S^{xx}_{120\degree} (N)$ close to the critical transition points along the $D_z$- and $J_2$-axes, respectively.
The critical strengths are found to be $D^c_z \sim 0.1$ at $J_2=0$, and $J^c_2\sim0.4$ at $D_z=0$.
We note that Fig. 1(c) shows a more apparent finite-size effect with an even-$N$/odd-$N$ alternation.
With additional $N=42$ calculations in the zero momentum sector, the critical strengths when extrapolated independently are $J^c_2(\textrm{even-}N)\sim 0.32$ and $J^c_2(\textrm{odd-}N)\sim 0.44$ at $D_z=0$.
$J^c_2$ is further reduced when $D_z$ is finite, and {\it vice versa}.

The $J_1$-$D_z$ and $J_1$-$J_2$ models have been separately investigated before.
In particular, $D^c_z=0.1$ at $J_2=0$ was also reported by previous ED studies~\cite{Cepas_PRB_2008,Rousochatzakis_PRB_2009}.
$J^c_2$ at $D_z=0$ has been computed by a number of methods~\cite{Tay_PRB_2011,Suttner_PRB_2014, Iqbal_PRB_2015, Gong_PRB_2015, Kolley_PRB_2015}, with the reported critical strength ranging from 0.2 to 0.7.
Classically, a positive infinitesimal $J_2$ would favour a $\mathbf{Q}=0$, 120\degree AFM long-range order ($J^c_2=0^+$)~\cite{Harris_PRB_1992, Spenke_PRB_2012}.
Our finding of $J^c_2\sim0.4$ highlights the role of quantum fluctuation in destabilizing magnetism.
Nonetheless, when $J_2$ becomes substantial, the quantum $\mathbf{Q}=0$, 120\degree AFM ground state can be stabilized.

{\it Materials Relevance --}
We next discuss the relevance of our phase diagram to different $S=1/2$, Cu-based materials with (nearly) perfect isotropic kagome structures, as denoted by the geometric symbols in Fig. 1(a).
In Cu(1,3-bdc), the material develops the $\mathbf{Q}=0$, 120\degree AFM order with a critical transition temperature $T_N\sim 2$ K~\cite{Nytko_JACS_2008,Marcipar_PRB_2009}.
Its interaction strengths $(D_z,J_2)\sim (0.1,0.05)$ estimated by first-principles calculations~\cite{Liu_arXiv_2014} indeed correspond to a positive $S^{xx}_{120\degree} \sim 0.05$ in our phase diagram.
The compound vesignieite also develops the $\mathbf{Q}=0$ magnetic order at $T_N\sim 9$ K~\cite{Colman_PRB_2011,Quilliam_PRB_2011,Yoshida_JMC_2012, Yoshida_JPSJ_2013}.
Its experimentally estimated DM interaction $\sim 0.15\, J_1$ renders a positive $S^{xx}_{120\degree} \sim 0.1$.
Similarly, Cs$_2$Cu$_3$SnF$_{12}$ is ordered at $T_N\sim 20$ K~\cite{Ono_PRB_2009,Matan_NatPhysics_2010, Katayama_arXiv_2014},
where the reported $D_z$ could be as large as $0.25\,J_1$, leading to $S^{xx}_{120\degree} \sim 0.15$.
Interestingly, a higher $T_N$ in experiments seems to be correlated with a larger positive $S^{xx}_{120\degree}$ in our phase diagram.

Herbertsmithite, however, is not magnetically ordered.
This could mean that its interactions $(D_z,J_2)$ are below the critical strengths $(D^c_z, J^c_2)\sim (0.1,0.4)$.
On the other hand, although $J_2$ is likely small compared to $J^c_2$, $D_z$ is reported to be comparable to $D^c_z$ in this compound.
For example, electron spin resonance experiment estimates a DM-interaction strength $\sim 0.08\,J_1$~\cite{Zorko_PRL_2008}.
Theoretical fit of $\chi_m$ indicates a $D_z \sim 0.1\, J_1$~\cite{Rigol_PRB_2007}.
In fact, first-principles calculations find $(D_z,J_2)\sim(0.1,0.05)$ for both Cu(1,3-bdc) and herbertsmithite~\cite{Liu_arXiv_2014},
whereas the former is magnetically ordered but the latter is not.
When the system resides in close proximity to the boundary of quantum phase transition,
additional perturbations such as spin-space exchange anisotropy~\cite{Ofer_PRB_2009, Han_PRL_2012, Kuroda_JPSJ_1995, Bekhechi_PRB_2003, Chernyshev_PRL_2014,Gotze_PRB_2015} or quenched dilution of Cu sites~\cite{Dommange_PRB_2003,Rigol_PRB_2007, Rozenberg_PRB_2008,Rousochatzakis_PRB_2009} in herbertsmithite could more easily suppress its long-range magnetism.

More recently, a structurally perfect kagome compound  ---the barlowite [Cu$_4$(OH)$_6$FBr]--- has been synthesized~\cite{Han_PRL_2014}. This material develops long-range magnetic order at $T_N=15$ K with a weak {\it ferromagnetic} moment.
In terms of our phase diagram, such a relatively high $T_N$ would imply a strong $D_z \sim 0.2\,J_1$,
which in conjunction with a small $D_\parallel$ can lead to a $\mathbf{Q}=0$, {\it canted} AFM state with a net ferromagnetic moment pointing outward from the kagome plane~\cite{Elhajal_PRB_2002, Harald_arXiv_2014}.
Future single-crystal measurements can further clarify the nature of barlowite's low-temperature magnetic structure.

{\it Gapped versus Gapless Spin Liquids --}
We next address the issue of spin gap by focusing on the $\Delta S^z_{\textrm{total}} = 1$ excitation:
$\Delta_T \equiv E_0(S^z_{\textrm{total}}=1) - E_0(S^z_{\textrm{total}}=0)$.
Here, $E_0(S^z_{\textrm{total}})$ is the lowest energy in a given $S^z_{\textrm{total}}$ sector,
and $\Delta_T$ corresponds to the triplet excitation when spin $SU(2)$ symmetry is present.
We will consider only even-$N$ clusters based on the gap definition.

\begin{figure}[t!]
\includegraphics[width=\columnwidth]{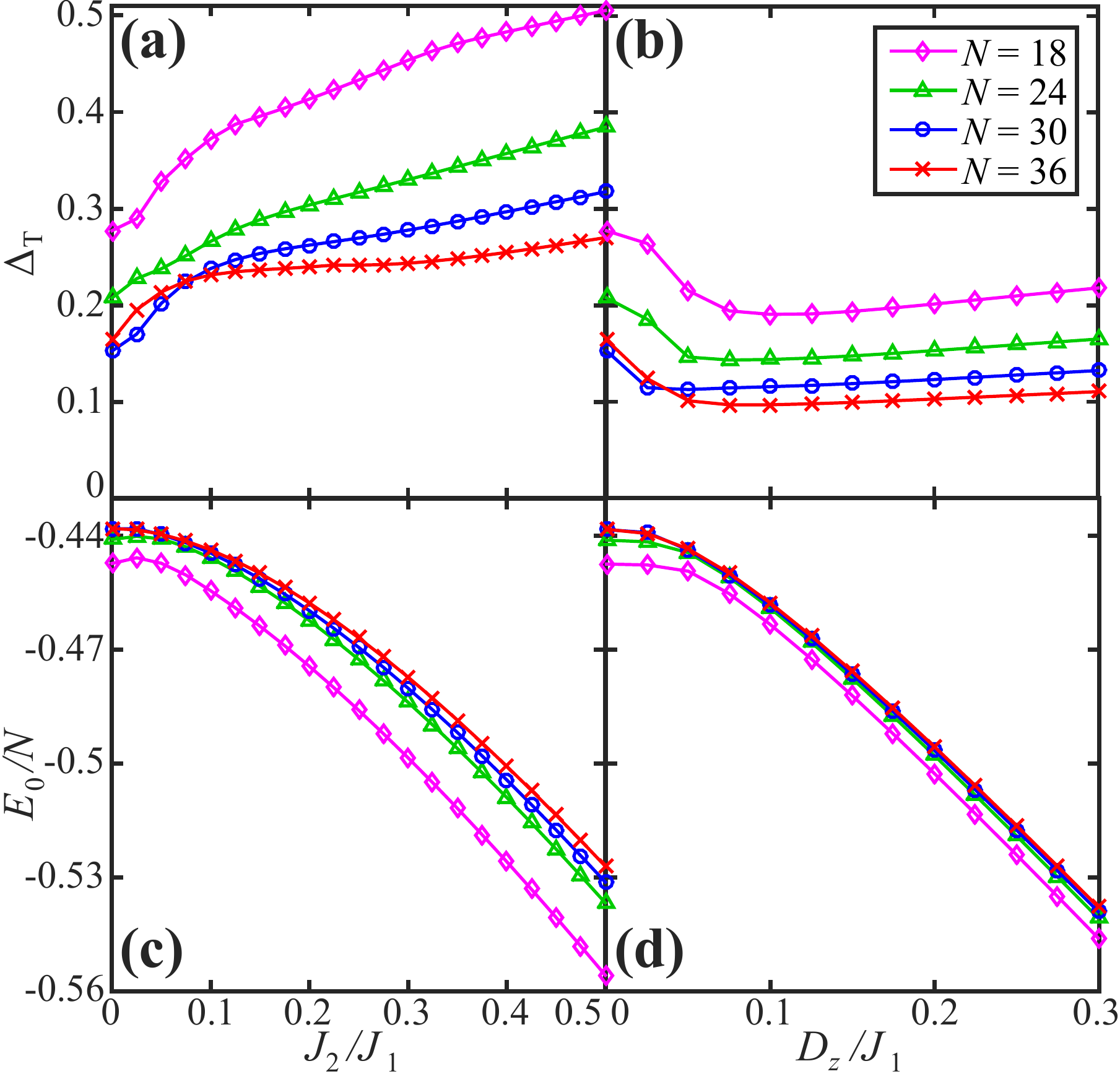}
\caption{
The spin gap $\Delta_T$ computed on clusters of size $N$ along (a) the $J_2$-axis ($D_z=0$) and (b) the $D_z$-axis ($J_2=0$).
The corresponding ground state energy per site $E_0/N$ is given in (c) and (d), respectively.
}
\end{figure}

We first note that extrapolating the gap on finite-size clusters is more difficult.
In the disordered region, the scaling form is {\it a priori} unknown.
The result can depend largely on the extrapolation function~\cite{Waldtmann_EPJB_1998, Lauchli_PRB_2011, Hiroki_JPSJ_2011},
as well as the cluster size and shape~\cite{Nishimoto_NatComm_2013}.
In the ordered phase, the energy splitting between quasi-degenerate ground states in quantum antiferromagnets would scale to zero as $1/N$~\cite{Anderson_PR_1952,Bernu_PRB_1992},
and the linear-dispersing Goldstone modes for spontaneous broken symmetries would scale as $1/\sqrt{N}$~\cite{Yildirim_PRB_2006, Messio_PRB_2010, Chernyshev_PRL_2014}.
But there may be no clear separation between these states in relatively small systems.
Despite these difficulties, however, the variation of the gap with parameters would be robust and distinguishable in our numerically exact data.
Therefore, instead of making precise quantitative statements, we would mainly focus on the trends.

Figure 2 shows the spin gap $\Delta_T$ and the ground state energy per site $E_0/N$ on different size clusters.
As shown in Fig. 2(a), $\Delta_T$ is quickly enhanced by $J_2$~\cite{Sindzingre_EPL_2009},
but the rate of increase becomes smaller above $J_2\sim0.1$ and tends to level off with increasing $N$.
If a simple $1/N$ scaling is employed in the disordered region as in previous ED study~\cite{Waldtmann_EPJB_1998},
$\Delta_T$ would reach its maximum at $J_2\sim 0.1$ and then decrease monotonically above it.
This extrapolated gap behaviour agrees well with recent density matrix renormalization group calculations~\cite{Kolley_PRB_2015}, implying that the gapped spin liquid phase is most stable around $J_2=0.1$.
At large $J_2$, although the raw data of $\Delta_T$ appear to grow with $J_2$ across the magnetic phase boundary [$J^c_2(\textrm{even-}N)\sim 0.32)$],
we note that the absolute value of the ground state energy $| E_0 |$ is also increasing [Fig. 2(c)].
The ratio $\Delta_T/ |E_0|$ becomes nearly flat and decreases systematically with increasing $N$ in the ordered regime.
The gap would eventually scale to zero in the thermodynamic limit, corresponding to spontaneous spin symmetry breaking of the ordered ground state.

The effect of $D_z$ on the spin gap is quite different.
As shown in Fig. 2(b), $\Delta_T$ is rapidly reduced in the presence of a small $D_z$~\cite{Mohakud_JPCS_2012}.
In fact, a simple $1/N$ extrapolation would indicate that $\Delta_T$ is already zero in the disordered region before reaching the critical point $D^c_z \sim 0.1$.
This shows the possibility of a gapless spin liquid ground state induced by spin exchange anisotropy~\cite{Ryu_PRB_2007,Hu_arXiv_2015}.
At $D_z \geq$~0.1, the ratio $\Delta_T/|E_0|$ stays flat and again decreases systematically with increasing $N$.
The thermodynamic-limit $\Delta_T$ remains gapless in the magnetic phase.

The above results show that small $J_2$ and $D_z$ could stabilize respectively a gapped and a gapless spin liquid.
Based on the magnetically disordered ground state and gapless excitations found in herbertsmithite,
our study suggests in this material a DM-interaction strength $0.05 \lesssim D_z \lesssim 0.1$, closely proximal to the quantum critical point $D^c_z$.
With the prevalent observation of gapless excitations in putative spin-liquid phases,
our study also implies that the DM interaction is in general non-negligible in isotropic kagome systems.

{\it Static Structure Factor --}
To connect with neutron scattering experiments, we study the static structure factor:
\begin{eqnarray}
S^{\alpha\beta} (\mathbf{Q}) \equiv \frac{1}{N} \sum_{ij} e^{i\mathbf{Q}\cdot(\mathbf{R}_i-\mathbf{R}_j)}\langle S^\alpha_i  S^\beta_j \rangle,
\end{eqnarray}
where $\alpha$, $\beta$ $\in \{x,y, z\}$.
Neutron scattering also provides energy-resolved spectra by measuring the dynamic structure factor $S^{\alpha\beta}(\mathbf{Q},\omega)$, where $S^{\alpha\beta}(\mathbf{Q}) = \int d\omega S^{\alpha\beta}(\mathbf{Q},\omega)$.
A spin liquid phase would produce continuous or diffusive scattering spectra, whereas an ordered magnet would generate sharp, discrete Bragg peaks.

\begin{figure}[t!]
\includegraphics[width=\columnwidth]{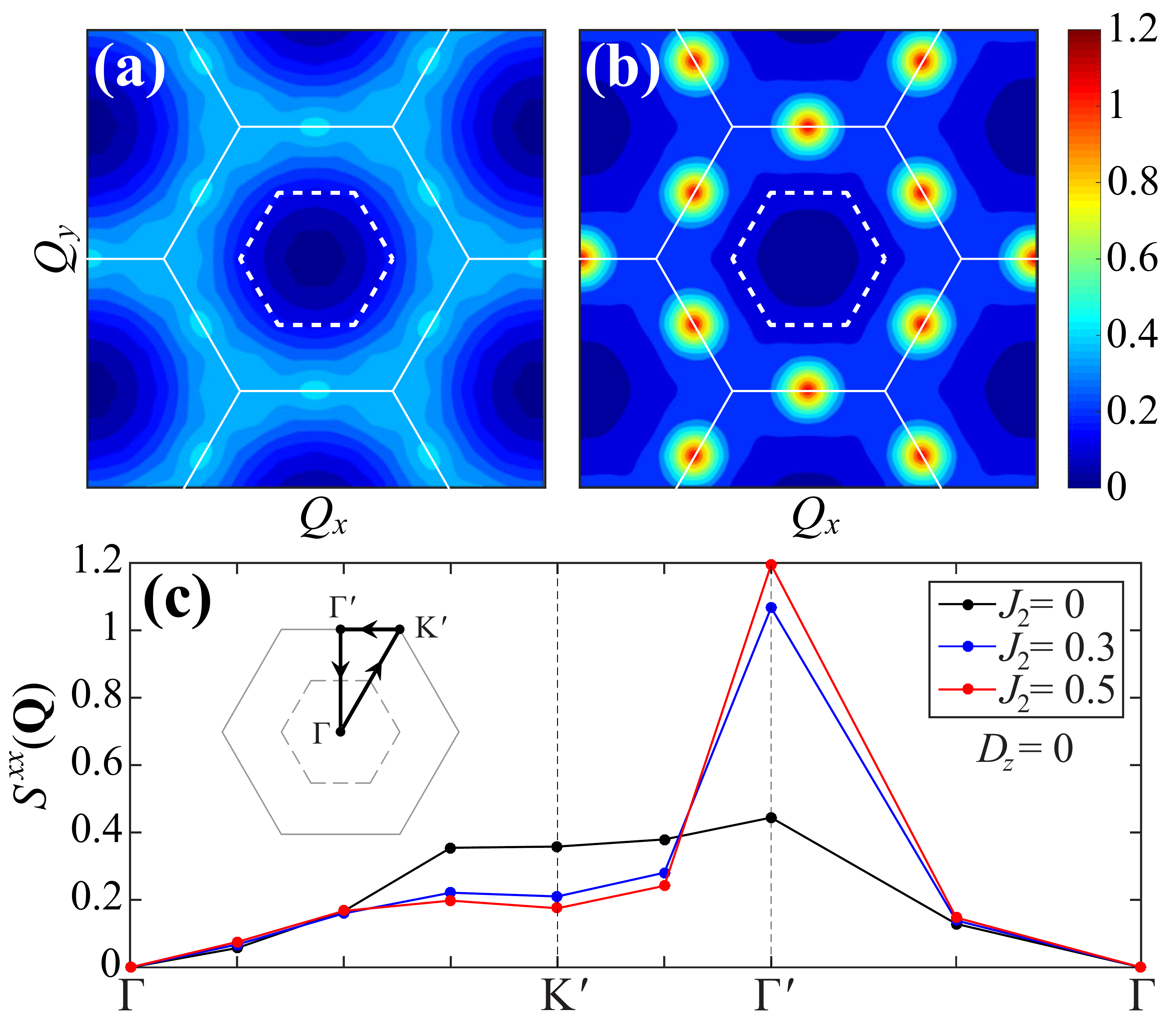}
\caption{
Transverse static structure factors $S^{xx}(\mathbf{Q})$ of the $J_1$-$J_2$-$D_z$ model for $J_2=0$ at (a) $D_z=0$ and (b) $D_z = 0.125$.
The first and extended Brillouin zones are indicated by the dashed and solid hexagons, respectively.
(c) $S^{xx}(\mathbf{Q})$ line cuts along high-symmetry points of the Brillouin zone at various $J_2$ with $D_z=0$.
}\label{fig:sofq}
\end{figure}

Figure \ref{fig:sofq} shows the transverse component $S^{xx}(\mathbf{Q})$ computed on the $N=36$ cluster.
When $D_z$ and $J_2$ are both zero [Fig. \ref{fig:sofq}(a)], the spectrum is close to being uniformly distributed along the extended Brillouin zone (BZ); the first BZ contains little spectral weight~\cite{Lauchli_arXiv_2009}.
This suggests that spin correlations are predominantly antiferromagnetic, while correlation lengths are on the order of lattice spacing.
Due to spin $SU(2)$ symmetry at $D_z=0$, the longitudinal component $S^{zz}(\mathbf{Q})$ is identical to $S^{xx}(\mathbf{Q})$,
and both components are zero at the $\Gamma$ point.

When $D_z=0.125$, the $\mathbf{Q}=0$, 120\degree AFM ground state manifests a structure factor that peaks at the midpoints of the extended BZ edges [Fig. \ref{fig:sofq}(b)].
In this case, $S^{zz}(\mathbf{Q})$ is much weaker than $S^{xx}(\mathbf{Q})$ at $D_z\ne 0$, and spins mainly lie in the $xy$-plane.
When the system is ordered, $S^{xx}(\mathbf{Q})$ at large $D_z$ and at large $J_2$ are in general similar,
except that (i) $S^{xx}(\mathbf{Q}=\Gamma)$ is no longer zero in the former, and (ii) $S^{xx}(\mathbf{Q}=K')$ is further suppressed in the latter.
In addition, the overall spectra do not undergo a sharp transition across the critical point $D^c_z$ or $J^c_2$.
These features can be seen in Fig. \ref{fig:sofq}(c) that shows high-symmetry line cuts of  $S^{xx}(\mathbf{Q})$ at various $J_2$ with $D_z=0$.

In herbertsmithite, neutron scattering signals are diffused for all the measured energies between 0.25 to 11 meV (where $J_1\sim 17$ meV)~\cite{Han_Nature_2012}.
The spectral weight is concentrated in the extend BZ but does not peak at any specific $\mathbf{Q}$ point,
although at 0.75 meV additional peak appears at the midpoints of the extended BZ edges.
These results agree with our $S^{xx}(\mathbf{Q})$ calculation for the magnetically disordered state [Fig. \ref{fig:sofq}(a)].
The experimental spectra also contain a small but finite weight at the $\Gamma$ point,
which could result from the DM interaction.
We note, however, that the experimental intensity integrated up to 11 meV contains only $\sim 20\%$ of the total spectral weight.
A more detailed theory-experiment comparison would require direct calculations of $S^{\alpha\beta}(\mathbf{Q},\omega)$.

In conclusion, we have studied the interplay between further-neighbour exchange and Dzyaloshinskii-Moriya interaction on the kagome lattice.
The phase diagram of the $J_1$-$J_2$-$D_z$ model is shown to contain various novel states of matter, including a $\mathbf{Q}$~=~0, 120\degree antiferromagnetic long-range order, as well as gapped and gapless quantum spin liquids.
A small variation of the parameters near the phase transition boundary could potentially account for the distinct properties observed in different kagome materials.
The phase diagram thereby serves as a benchmark for determining the importance of these additional perturbations.
Studying the dynamical properties in different parts of the phase diagram and making further connection to inelastic neutron or x-ray scattering measurements would be important for future research.


\begin{acknowledgments}
The authors acknowledge discussions with Keun Hyuk Ahn, Hong-Chen Jiang, and Zhenyue Zhu.
C.C.C. is supported by the Aneesur Rahman Postdoctoral Fellowship at Argonne National Laboratory, operated by the U.S. Department of Energy (DOE) Contract No. DE-AC02-06CH11357.
R.R.P.S. is supported by the National Science Foundation Grant No. DMR-1306048.
T.F.S. and M.v.V are supported by the U.S. DOE, Office of Basic Energy Sciences, under Award No. DE-FG02-03ER46097, and by the Institute for Nanoscience, Engineering and Technology at Northern Illinois University.
This research used resources of the National Energy Research Scientific Computing Center, supported by the U.S. DOE under Contract No. DE-AC02-05CH11231.
\end{acknowledgments}

\bibliography{kagome}

\clearpage
\begin{figure}[p]
\vspace{-0.75in}\hbox{\hspace{-0.75in}\includegraphics{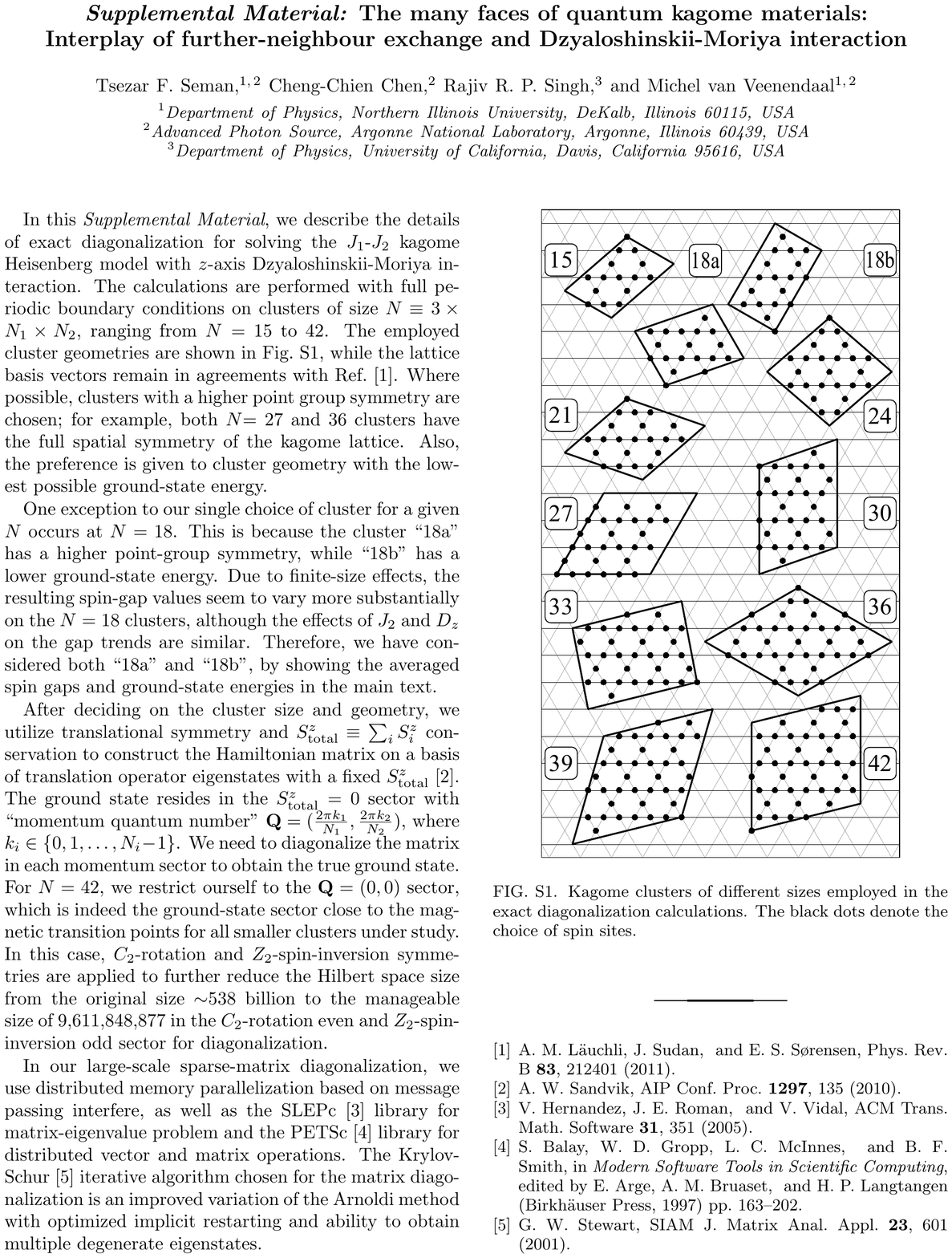}}    
\end{figure}
\thispagestyle{empty}

\end{document}